\documentclass[twocolumn,preprintnumbers,bibnotes,amsmath,amssymbaps,pra,superscriptaddress,longbibliography]{revtex4-2}
\usepackage[version=3]{mhchem}
\usepackage[dvipsnames]{xcolor}
\usepackage[colorlinks=false, citecolor=blue, urlcolor=blue, linkcolor=blue]{hyperref}
\usepackage[full]{textcomp}
\usepackage{graphicx}
\usepackage{amsmath}
\usepackage{fixmath}
\usepackage{amsfonts}
\usepackage{amssymb}
\usepackage{braket}
\usepackage{subfigure}
\usepackage{physics}
\usepackage{stix}
\usepackage{mathtools}
\DeclareUnicodeCharacter{0360}{\eth}
\newcommand{\AK}[1]{{#1}}

\begin{document}

\title{Evolution of a twisted electron wave packet perturbed by an inhomogeneous electric field}

\author{A.~Kudlis}
\email{andrewkudlis@gmail.com}
\affiliation{Science Institute, University of Iceland, Dunhagi 3, IS-107 Reykjavik, Iceland}
\author{I.~A.~Aleksandrov}
\email{i.aleksandrov@spbu.ru}
\affiliation{Department of Physics, Saint Petersburg State University, Universitetskaya Naberezhnaya 7/9, Saint Petersburg 199034, Russia}
\affiliation{Ioffe Institute, Politekhnicheskaya Street 26, Saint Petersburg 194021, Russia}
\author{N.~N.~Rosanov}
\affiliation{Ioffe Institute, Politekhnicheskaya Street 26, Saint Petersburg 194021, Russia}


\begin{abstract}
Laguerre-Gaussian (LG) wave packets, known for their vortex structure and nonzero orbital angular momentum (OAM), are of great interest in various scientific fields. Here we study the nonrelativistic dynamics of a spatially-localized electron LG wave packet interacting with an inhomogeneous external electric field that violates the axial symmetry of the initial wave function. We focus on the analysis of the electron density and demonstrate how it is affected by the external field. Within the first order of perturbation theory, we calculate the electron wave function and reveal that the electric field may significantly alter the wave packet's structure and distort its qualitative form. We demonstrate that due to the interaction with the external field, the degenerate zeros of the initial wave function located on the $z$ axis split into multiple nondegenerate nodes in the transverse plane representing separate single-charge vortices. This mechanism resembles the analogous effects known in topological optics. These findings provide new insights into controlling and manipulating twisted matter beams and into their possible instabilities.
\end{abstract}

\maketitle


\section{Introduction} \label{sec:intro}

Singular optics and its generalization, topological optics~\cite{Soskin2001,BerryDennis2001,Freund2002,RuchiSenthilkumaranPal2020a,Simon2021}, which involve the analysis of properties and transformations of sets of points where various characteristics of optical radiation cannot be defined, have found numerous applications in the formation of optical traps, manipulation of microparticle positions, laser beam transport, and encoding of information carried by such beams~\cite{GahaganSchwartzlander1996,FrieseNieminenHeckenbergRubinszteinDunlop1998,ONeilPadgett2000,RuchiSenthilkumaran2020b,AngelskyBekshaevHansonZenkovaMokhunJun2020,WangTuLiWang2021}. An important example of such singularity is optical vortices. At their center, the intensity of radiation falls to zero and, accordingly, the phase is undefined. Moreover, upon encircling the center along a closed contour in the plane orthogonal to the main propagation direction of the beam, the wave phase changes by $\pm 2\pi m$, where $\pm m$ ($m = 0, 1, 2, \dots$) is the integer {\it topological charge}.

A common example of a vortex beam is the optical Laguerre-Gaussian (LG) beams characteristic of laser radiation~\cite{Siegman1986}. It is well known that in a homogeneous linear medium, singularities with higher topological charges of these and other vortex optical beams split into singularities with unit charges ($m = 1$) under perturbations~\cite{BaranovaZeldovich1981}. The same splitting occurs in a homogeneous medium with Kerr nonlinearity~\cite{KruglovLogvinVolkov1992}. At the same time, in a homogeneous medium with dissipative nonlinearity, vortex beams with higher charges can be stable~\cite{FedorovRosanovShatsevVeretenovVladimirov2003}. Deep analogies between optics and quantum mechanics allow many concepts and conclusions to be effectively transferred between these branches of physics. In the present investigation, we examine the stability and dynamics of a vortex (twisted) {\it electron} wave packet interacting with an inhomogeneous electric field. Our goal is to explore how external perturbations affect the vortex structure of the initial LG wave packet. In particular, we will see whether the singularities of the electron wave function split due to the interaction, i.e., we will compare the electron dynamics with the above mentioned mechanisms revealed in optical setups. 

The study of LG optical beams, characterized by their unique vortex structure and ability to carry a nonzero orbital angular momentum (OAM), has garnered significant interest in quantum optics, accelerator physics, and other fields of study~\cite{PhysRevA.45.8185,BEIJERSBERGEN1993123,sasaki2008proposal,Verbeeck2010,jentschura2011generation,katoh2017angular,chen2018gamma,Karlovets2021,IVANOV2022103987,PhysRevA.110.032202,Begin2025}. These modes offer a rich framework for exploring light-matter interactions and potential applications in optical tweezers, quantum information processing, and high-resolution microscopy~\cite{Torres2011-uf,Padgett2011} (see also reviews~\cite{yao2011orbital,knyazev2018beams,erhard2018twisted,Shen2019}, where various fundamental aspects of OAM in light are discussed).

Although the field initially focused on the study of photons, it soon became evident that twisted states of massive particles could be generated as well. It was found that not only photons but also particles like electrons could carry OAM~\cite{LearyNJP2008,BliokhPRL2011,BLIOKH20171,RevModPhys.89.035004}). \AK{The structure of the wave function of a twisted (vortex) electron is similar to that of a twisted photon and possesses rotational symmetry with respect to the mean propagation direction coinciding with, say, the $z$ axis.} A notable aspect of twisted electrons for practical applications is that, in contrast to photons, they have a magnetic moment directly proportional to their OAM, which can experimentally reach values in the hundreds of $\hbar$~\cite{Mafakheri_2017}. 
Subsequent investigations have shown that vortex electrons are sensitive to various external fields~\cite{silenko2017manipulating,aleksandrov_pra_2022,PhysRevA.110.052207,PhysRevA.109.L040201} (see also Refs.~\cite{ivanov2011colliding,maruyama2019compton,Neha2024,fan_prd_2025} for the discussion of various setups involving twisted particles).

\begin{figure}[t!]
    \centering    \includegraphics[width=0.92\linewidth]{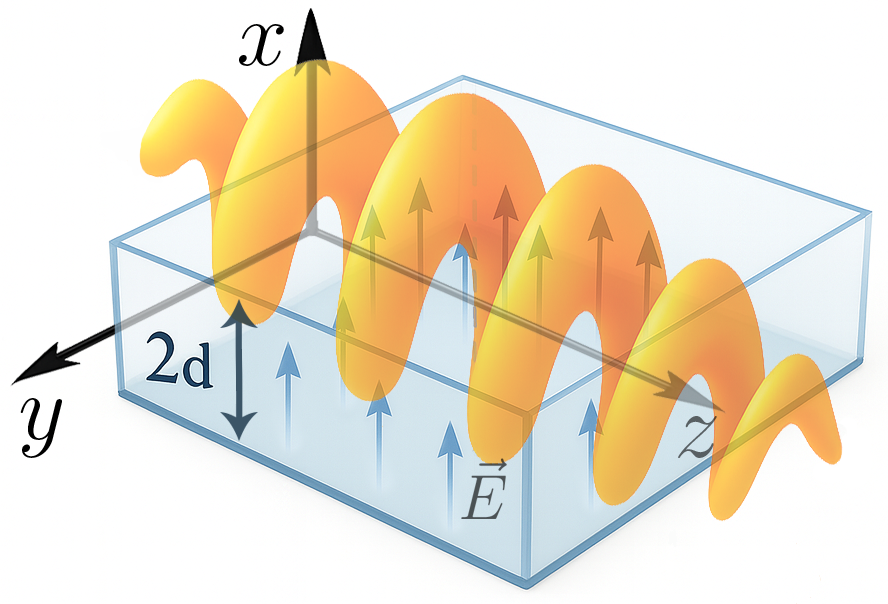}
\caption{\AK{Simplified scheme of the main setup under consideration. A twisted electron wave packet with a given $z$ projection $l$ of orbital angular momentum evolves under the influence of an external electric field directed along the $x$ axis. The electric field is present in a finite spatial layer $2d$ and has a smooth $x$-dependent profile. The external field breaks the initial rotational symmetry of the vortex state. By means of first-order perturbation theory, we investigate the dynamics of the wave packet and examine its topological properties.}}
    \label{fig:scheme}
\end{figure}

Despite the advances in generating and manipulating twisted electron states, a gap remains in understanding how external perturbations that violate the cylindrical symmetry affect the stability of such beams. In this paper, we analyze an electron LG wave packet with nonzero OAM that interacts with a static inhomogeneous electric field. As a specific examples, we consider a fully-localized external field described by an off-axis (with respect to $z$) delta-function and an electric field oriented along the transverse $x$ axis \AK{(the latter setup is illustrated in Fig.~\ref{fig:scheme})}. These configurations intentionally break the rotational symmetry about the initial beam axis $z$ and thus provide a simple models for examining how external perturbations affect the stability of vortex electrons. More specifically, we investigate the time evolution of the electron density, focusing on cross-sectional profiles of the wave packet. By employing a perturbative approach, we derive the first-order corrections to the wave function and evaluate how the probability density profile deforms under the influence of the external potential. Our results uncover notable OAM mixing and splitting of the central minimum in the electron density distribution, illustrating the importance of small external perturbations.  These findings advance our fundamental understanding of structured beams and set the stage for future experimental and theoretical studies aimed at controlling and exploiting vortex electrons in real applications.

The paper is organized as follows. In Sec.~\ref{sec:setup}, we introduce the basic setup and describe the system under consideration. In Sec.~\ref{sec:PT}, we outline the theoretical method and develop a perturbative framework used to analyze the interaction of a twisted wave packet with an external electric field. Next, in Sec.~\ref{sec:results}, we present and discuss our numerical results, highlighting the key effects of the inhomogeneous electric field. Finally, in Sec.~\ref{sec:conclusion}, we provide a conclusion.

\allowdisplaybreaks
\section{Setup} \label{sec:setup}

We assume that the initial nonrelativistic twisted wave packet at $t=0$ is described by the following wave function in coordinate space:
\begin{equation}
\Psi_0(\mathbf{r}) = \frac{\sigma^{3/2}}{\pi^{3/4}\sqrt{l!}} \, (\sigma\rho)^l \mathrm{e}^{i \bar{p} z/\hbar} \, \mathrm{e}^{il\varphi} \, \mathrm{exp} \bigg [-\frac{\sigma^2}{2} \, (z^2 + \rho^2) \bigg ],
\label{eq:psi0}
\end{equation}
where $\rho$ and $\varphi$ are the polar coordinates in the $xy$ plane. Here $\sigma$ is the characteristic width of the wave packet in momentum space, $\bar{p}$ is the central momentum along the $z$ axis, and $l$ is the $z$ projection of the electron OAM in units of $\hbar$ (we assume $l>0$). The wave function~\eqref{eq:psi0} is localized in all of the three spatial directions and has a unit $L^2$ norm. A more general form of a Laguerre-Gaussian wave packet can involve associated Laguerre polynomials $L_n^{|l|}$~\cite{Karlovets2021}, but here we will focus on $n=0$ and positive $l$.

Consider the interaction of the electron wave packet with an inhomogeneous external electric field described by a static scalar potential $\phi (\mathbf{r})$. The full time‐dependent Schr\"odinger equation reads
\begin{equation}
i \hbar \frac{\partial}{\partial t}\,\Psi(t,\mathbf{r}) = \left[-\frac{\hbar^2}{2m}\Delta + e \phi(\mathbf{r})\right] \Psi(t,\mathbf{r}),
\end{equation}
where $m$ and $e$ are the electron mass and charge, respectively. The initial condition is given by $\Psi(0,\mathbf{r})=\Psi_0(\mathbf{r})$. Since we are interested in investigating rather weak perturbations, our main theoretical tool will be perturbation theory with respect to the external potential $\phi(\mathbf{r})$. Let us rewrite the Schr\"odinger equation in the following form:
\begin{equation}
\left( i \hbar \frac{\partial}{\partial t} + \frac{\hbar^2}{2m}\Delta \right ) \Psi(t,\mathbf{r}) = e \phi(\mathbf{r}) \Psi(t,\mathbf{r}).
\end{equation}
To solve this equation by means of perturbation theory, let us introduce the Green's function,
\begin{equation}
\left( i \hbar \frac{\partial}{\partial t} + \frac{\hbar^2}{2m}\Delta \right ) G(t,\mathbf{r}) = i \hbar \delta (t) \delta (\mathbf{r}).
\end{equation}
This Green's function is known in an explicit form,
\begin{equation}
G (t, \mathbf{r}) = \theta(t) \bigg ( \frac{m}{2\pi i \hbar t} \bigg)^{3/2} \, \mathrm{exp} \bigg ( \frac{im \mathbf{r}^2}{2\hbar t}\bigg ).
\end{equation}
The first-order correction can then be calculated via
\begin{equation}
\Psi^{(1)}(t,\mathbf{r}) = -\frac{i}{\hbar} \int \limits_0^{\infty} \! dt' \int \! d\mathbf{r}' \, G(t - t',\mathbf{r} - \mathbf{r}') e \phi(\mathbf{r}') \Psi^{(0)}(t',\mathbf{r}'),
\label{eq:PT1_G}
\end{equation}
\AK{where $\Psi^{(0)}(t,\mathbf{r})$ denotes the unperturbed wave packet, which incorporates the usual free-electron spreading effects.}

Next, let us specify the external electric field that provides a perturbation of the electron wave packet. We will consider two different setups. First, we will discuss a fully-localized external field in the following very simple form:
\begin{equation}
\phi (\mathbf{r}) = \lambda \delta (\mathbf{r} - \mathbf{r}_0).
\label{eq:app_pot}
\end{equation}
This model describes a static external field localized at the position $\mathbf{r}_0$.

Second, we will consider an external electric field directed along the $x$ axis, so that it violates the axial symmetry of the initial state. It turns out that a homogeneous time-dependent external field $\mathbf{E} = E(t) \mathbf{e}_x$ does not affect the structure of the electron state and only provide an acceleration of the whole wave packet (we prove this fact in the Appendix). In order to analyze a nontrivial space-dependent perturbation, we assume that the external field has a static localized profile in the transverse $x$ direction and is described by the following scalar potential:
\begin{equation}
\phi (\mathbf{r}) = \phi(x) =- E_0 d f \bigg (\frac{x - a}{d}\bigg),
\label{eqn:profile}
\end{equation}
where $E_0$ is the electric field strength, $a$ is its central position, and $d$ is its spatial width. The dimensionless function $f(\xi)$ obeys $f(\xi<-1) = -1$ and $f(\xi>1) = 1$. It can be represented as
\begin{equation}
f(\xi) = 2 \theta (\xi) - 1 + g(\xi),
\label{eq:fg}
\end{equation}
where $\theta (\xi)$ is a Heaviside step function and $g(\xi)$ vanishes for $|\xi| > 1$. In what follows, the representation~\eqref{eq:fg} will be particularly useful within our analytical treatment of the problem in momentum space.

The field configuration~\eqref{eqn:profile} is considered as a toy model as it is sufficiently simple to perform analytical and numerical calculations. However, an $x$-dependent potential can be implemented, for example, with a pair of parallel plate electrodes whose length greatly exceeds the transverse size of the electron beam and whose separation varies slowly along $z$. By shaping the electrode edges (using razor-blade masks or lithographically patterned conductors) one prescribes the voltage drop $V(x)$ and hence the static field $E_x(x)=-dV/dx$.  Because the plates extend uniformly along $y$ and parallel to the beam axis $z$, the resulting field is effectively homogeneous in those directions within the interaction region, matching the one-dimensional profile adopted in Eq.~\eqref{eqn:profile}. The essential feature of this setup is the fact that it violates the axial symmetry of the initial state~\eqref{eq:psi0}. The effects of this violation are our main focus. 

In what follows, we will analyze the evolution of the twisted wave packet taking into account the leading-order contribution of the interaction term $e\phi(\mathbf{r})$. We will be primarily interested in how the structure of the electron wave function is modified due to the external perturbation.


\section{Theoretical description} \label{sec:PT}

Here we will outline our theoretical approach based on the perturbative expansion with respect to the external electric field and perform necessary calculations for the two specific setups described above.

\subsection{Zeroth-order solution} \label{sec:Psi0}

If one disregards the external potential $\phi (\mathbf{r})$, then the Schr\"odinger equation with the initial condition~\eqref{eq:psi0} can be solved analytically. This can be done, for instance, in the momentum representation, where the initial wave function is given by
\begin{align}
\Phi_0 (\mathbf{p}) &\equiv \int \! d\mathbf{r} \mathrm{e}^{-\frac{i}{\hbar}\mathbf{p}\mathbf{r}}\Psi_0 (\mathbf{r}) = \frac{2^{3/2} \pi^{3/4}}{i^l \sqrt{l!}}\frac{1}{\sigma^{3/2}}\left(\frac{p_{\perp}}{\sigma\hbar}\right)^l \mathrm{e}^{il\varphi_{p}} \nonumber \\
{}&\times \mathrm{exp} \bigg [ -\frac{(p_z-\bar{p})^2}{2\sigma^2\hbar^2} -\frac{p_{\perp}^2}{2\sigma^2\hbar^2} \bigg ].
\label{eq:Phi0_explicit}
\end{align}
Here the polar radius and angle of the momentum vector are denoted by $p_\perp$ and $\varphi_p$, respectively. The wave function $\Psi (t, \mathbf{r})$ can be represented in the following form:
\begin{equation}
\Psi(t,\mathbf{r})
=
\int \! \frac{d\mathbf{p}}{(2\pi\hbar)^3}\,
  \mathrm{e}^{\frac{i}{\hbar} \mathbf{p} \mathbf{r}} \mathrm{e}^{-\frac{i\mathbf{p}^2}{2m\hbar} t}
  \, \Phi (t, \mathbf{p}).
\label{eq:Psi_fourier}
\end{equation}
The zeroth-order contribution to the wave function $\Phi(t,\mathbf{p})$ in the momentum representation is stationary and reads
\begin{equation}
\Phi^{(0)} (t,\mathbf{p}) = \Phi^{(0)} (0,\mathbf{p}) = \Phi_0 (\mathbf{p}).
\label{eq:Phi0}
\end{equation}
This obviously describes the ordinary spreading of the wave packet in coordinate space:
\begin{align}
\Psi^{(0)} (t, \mathbf{r}) &= \frac{1}{\pi^{3/4}\sqrt{l!}} \, \frac{\rho^l}{[\sigma_\perp (t)]^{l+3/2}} \, \mathrm{exp} \bigg ( -\frac{i\bar{p}^2 t}{2m\hbar} \bigg ) \mathrm{e}^{i \bar{p} z/\hbar} \, \mathrm{e}^{il\varphi} \nonumber\\
{}&\times \mathrm{exp} \bigg \{ \! -i(l+3/2) \arctan (t/t_\text{d}) \nonumber \\
{}&- \frac{1-it/t_\text{d}}{2[\sigma_\perp (t)]^2} \big [ \rho^2 + (z - \bar{p} t/m)^2 \big ] \bigg \},
\label{eq:Psi_0}
\end{align}
where $t_\text{d} = m/(\sigma^2 \hbar)$ and $\sigma_\perp (t) = (1/\sigma) \sqrt{1+t^2/t_\text{d}^2}$.

\subsection{Fully-localized external field}

In the simplest case involving the external field in the form~\eqref{eq:app_pot}, the calculations can be performed immediately. According to Eq.~\eqref{eq:PT1_G}, the first-order correction can be calculated via
\begin{equation}
\Psi^{(1)}(t,\mathbf{r}) = -\frac{ie \lambda}{\hbar} \int \limits_0^{t} \! dt' \, G(t - t', \mathbf{r} - \mathbf{r}_0) \Psi^{(0)}(t',\mathbf{r}_0).
\end{equation}
By taking into account Eq.~\eqref{eq:Psi_0}, we obtain
\begin{align}
\Psi^{(1)}(t,\mathbf{r}) =& -\frac{ie\lambda/\hbar}{\pi^{3/4} \sqrt{l!}} \, \bigg ( \frac{m}{2\pi i \hbar} \bigg )^{3/2} \rho_0^{l} \, \mathrm{e}^{i \bar{p} z_0/\hbar} \, \mathrm{e}^{il\varphi_0} \nonumber \\
{}&\times \int \limits_0^{t} \! \frac{dt'}{(t - t')^{3/2}} \, \mathrm{exp} \bigg [ \frac{im (\mathbf{r} - \mathbf{r}_0)^2}{2\hbar (t - t')} - \frac{i\bar{p}^2 t'}{2m\hbar} \bigg ] \nonumber \\
{}&\times \mathrm{exp} \bigg \{ \! -i(l+3/2) \arctan (t'/t_\text{d}) \nonumber \\
{}&- \frac{1-it'/t_\text{d}}{2[\sigma_\perp (t')]^2} \big [ \rho_0^2 + (z_0 - \bar{p} t'/m)^2 \big ] \bigg \},
\label{eq:app_Psi_1}
\end{align}
where $\rho_0$, $\varphi_0$, and $z_0$ are cylindrical coordinates of $\mathbf{r}_0$. Without loss of generality, we assume $\varphi_0 = 0$. The azimuthal dependence of this correction is governed by the term $-2\rho \rho_0 \cos \varphi$ in $(\mathbf{r} - \mathbf{r}_0)^2$ in the second line of Eq.~\eqref{eq:app_Psi_1}. We immediately observe that in the vicinity of the origin $\rho = 0$, this term is suppressed, so $\Psi^{(1)}(t,\mathbf{r})$ is approximately independent of $\varphi$, i.e., it represents a vortex contribution with a zero OAM. The numerical analysis of the solution~\eqref{eq:app_Psi_1} will be carried out in Sec.~\ref{sec:results}.

\subsection{Electric field along the $x$ axis}

Here we will turn to the discussion of a more involved field configuration given by Eq.~\eqref{eqn:profile}. Although one can directly employ the prescription~\eqref{eq:PT1_G}, we will proceed in a different way by considering the problem in the momentum representation.

As was already seen in Sec.~\ref{sec:Psi0}, considering the evolution of the wave packet in momentum space is advantageous due to the very simple form of the unperturbed Hamiltonian in this representation. Let us first define the one-dimensional Fourier transform of the external potential:
\begin{align}
\hat{\phi} (q) &\equiv \int\limits_{-\infty}^{\infty} \! dx \, \mathrm{e}^{-iqx/\hbar} \phi(x) \nonumber \\
{}&=-E_0 d^2 \mathrm{e}^{-iqa/\hbar} \int\limits_{-\infty}^{\infty} \! d\xi \, \mathrm{e}^{-iqd \xi/\hbar} f(\xi).
\end{align}
The Fourier transform in the second line contains a generalized function arising from $2\theta (\xi) - 1$ in Eq.~\eqref{eq:fg} and an ordinary function that represents the Fourier transform of $g(\xi)$. Namely,
\begin{equation}
\hat{\phi} (q) = -E_0 d^2 \mathrm{e}^{-iqa/\hbar} \bigg [ \hat{g} (qd/\hbar) - \frac{2i\hbar}{d} \mathcal{P} \, \frac{1}{q} \bigg].
\label{eq:phi_P}
\end{equation}
Here $\mathcal{P}$ corresponds to the principal value integral and
\begin{equation}
\hat{g} (\eta) \equiv \int \limits_{-1}^{1} \! d\xi \, \mathrm{e}^{-i \eta \xi} g(\xi).
\end{equation}
The function $g (\xi)$ can be chosen in different ways and will be specified below.

The full time-dependent wave function $\Psi(t,\mathbf{r})$ is represented in the form~\eqref{eq:Psi_fourier}. The time-dependent exponential factor is introduced in order to simplify the corresponding equation for $\Phi (t, \mathbf{p})$ (this is nothing but a transition to the interaction picture):
\begin{multline}
  i \hbar \frac{\partial}{\partial t} \Phi(t,\mathbf{p})  = \int\limits_{-\infty}^{\infty} \! \frac{d p_x'}{2\pi\hbar} \, e\hat{\phi}(p_x-p'_x)\\ 
  \times \mathrm{e}^{\frac{it}{2m\hbar}(p_x^2-p_x'^2)} \, \Phi(t,p'_x,p_y,p_z).
\label{eq:Phi_eq}
\end{multline}
The zeroth-order contribution~\eqref{eq:Phi0} is explicitly given by Eq.~\eqref{eq:Phi0_explicit}. The first-order correction to the wave function in momentum space has now the following form:
\begin{multline}
\Phi^{(1)} (t,\mathbf{p})  = -\frac{i}{\hbar} \int \limits_0^t \! dt' \int\limits_{-\infty}^{\infty} \! \frac{d p_x'}{2\pi\hbar} \, e\hat{\phi}(p_x-p'_x)\\ 
  \times \mathrm{e}^{\frac{it'}{2m\hbar}(p_x^2-p_x'^2)} \, \Phi_0 (p'_x,p_y,p_z).
\label{eq:Phi_1}
\end{multline}
Although the integral over $t'$ can be evaluated immediately, we will retain this integral. Since we are interested in computing the position-space wave function, let us directly analyze the following first-order contribution:
\begin{equation}
\Psi^{(1)}(t,\mathbf{r}) = \int \! \frac{d\mathbf{p}}{(2\pi\hbar)^3}\,
  \mathrm{e}^{\frac{i}{\hbar} \mathbf{p} \mathbf{r}} \mathrm{e}^{-\frac{i\mathbf{p}^2}{2m\hbar} t}
  \, \Phi^{(1)} (t, \mathbf{p}).
\label{eq:Psi_1}
\end{equation}
We first substitute Eq.~\eqref{eq:Phi_1} into Eq.~\eqref{eq:Psi_1}, interchange the integration variables $p_x$ and $p_x'$, and then introduce dimensionless $\xi = (p_x' - p_x)d/\hbar$ instead of $p_x'$:
\begin{widetext}
\begin{equation}
\Psi^{(1)}(t,\mathbf{r}) = -\frac{i}{2\pi\hbar d} \int \limits_0^t \! dt' \int\limits_{-\infty}^{\infty} \! d\xi \int \! \frac{d\mathbf{p}}{(2\pi\hbar)^3}\, \mathrm{e}^{\frac{i}{\hbar} \mathbf{p} \mathbf{r}} \mathrm{e}^{i\xi x /d} \mathrm{e}^{-\frac{i\mathbf{p}^2}{2m\hbar} t} \mathrm{e}^{-\frac{i\xi}{2md} \, (2p_x + \hbar \xi/d)(t-t')} e\hat{\phi}(\hbar \xi /d) \Phi_0 (\mathbf{p}).
\end{equation}
Next, we will analytically calculate the three-dimensional integral over $\mathbf{p}$. The integral over $p_z$ reads
\begin{multline}
\int\limits_{-\infty}^{\infty} \! \frac{d p_z}{2\pi\hbar} \, \mathrm{e}^{\frac{i}{\hbar} p_z z} \mathrm{e}^{-\frac{ip_z^2}{2m\hbar} t} \mathrm{exp} \bigg [ -\frac{(p_z-\bar{p})^2}{2\sigma^2\hbar^2} \bigg ] = \frac{1}{\sqrt{2\pi}} \sqrt{\frac{\sigma}{\sigma_\perp (t)}} \, \mathrm{e}^{\frac{i}{\hbar} \bar{p} z} \mathrm{e}^{-\frac{i\bar{p}^2}{2m\hbar} t} \\
{}\times \mathrm{exp} \bigg \{ \! -(i/2) \arctan (t/t_\text{d}) - \frac{1-it/t_\text{d}}{2[\sigma_\perp (t)]^2} \, (z - \bar{p} t/m)^2 \bigg \} \equiv \frac{\sigma}{\sqrt{2\pi}} \, Q(t,z),
\label{eq:Q}
\end{multline}
It is also possible to directly perform the $p_x$ integration:
\begin{equation}
\int\limits_{-\infty}^{\infty} \! \frac{d p_x}{2\pi\hbar} \, \mathrm{e}^{\frac{i}{\hbar} p_x x} \mathrm{e}^{-\frac{ip_x^2}{2m\hbar} t} \mathrm{e}^{-\frac{i\xi p_x}{md} \, (t-t')} \mathrm{e}^{-\frac{p_x^2}{2 \sigma^2 \hbar^2}} \, (p_x + ip_y)^l = \frac{i^l}{2\pi \hbar} \sqrt{\frac{\pi}{\alpha}} \frac{\mathrm{e}^{-\beta^2/(4\alpha)}}{\alpha^l} \, F_l \bigg (\alpha, p_y + \frac{\beta}{2\alpha} \bigg),
\end{equation}
where
\begin{align}
\alpha &\equiv \frac{1+it/t_\text{d}}{2\sigma^2 \hbar^2}, \qquad \beta \equiv \frac{x}{\hbar} - \frac{\xi (t - t')}{md}, \label{eq:alpha_beta} \\
F_l (\alpha, w) &\equiv \sum_{n=0}^{[l/2]} (-1)^n C_l^{2n} \frac{(2n-1)!!}{2^n} \, \alpha^{l-n} w^{l-2n}.
\end{align}
Note that $\beta$ here is a function of $\xi$ (and also $t$ and $x$), and $p_y$ is involved in the argument of $F_l$. Next, let us calculate the following integral with a monomial $[p_y + \beta/(2\alpha)]^k$:
\begin{align}
\int\limits_{-\infty}^{\infty} \! \frac{d p_y}{2\pi\hbar} \, \mathrm{e}^{\frac{i}{\hbar} p_y y} \mathrm{e}^{-\frac{ip_y^2}{2m\hbar} t} \mathrm{e}^{-\frac{p_y^2}{2 \sigma^2 \hbar^2}} \, \bigg (p_y + \frac{\beta}{2\alpha} \bigg)^k &= \frac{1}{2\pi \hbar} \sqrt{\frac{\pi}{\alpha}} \frac{\mathrm{e}^{-y^2/(4\alpha \hbar^2)}}{\alpha^k} \, G_k \bigg (\alpha, \frac{iy}{2\alpha \hbar} + \frac{\beta}{2\alpha} \bigg), \\
G_k (\alpha, u) &\equiv \sum_{n=0}^{[k/2]} C_k^{2n} \frac{(2n-1)!!}{2^n} \, \alpha^{k-n} u^{k-2n}.
\end{align}
By combining our results, we obtain
\begin{align}
\Psi^{(1)}(t,\mathbf{r}) &= -\frac{i}{2\hbar d} \frac{1}{\pi^{7/4} \sqrt{l!}} \frac{\sigma^{3/2}}{[\sigma \sigma_\perp(t)]^{2l+2}} \, \bigg ( 1 - \frac{it}{t_\text{d}} \bigg )^{l+1} Q(t,z) \, \mathrm{exp} \bigg [ \! - \frac{y^2 ( 1 - it/t_\text{d} )}{2[\sigma_\perp(t)]^2} \bigg ] \nonumber \\
{}&\times \int \limits_0^t \! dt' \int\limits_{-\infty}^{\infty} \! d\xi \, \Xi_l (t, t', x, y, \xi) e\hat{\phi}(\hbar \xi /d) \mathrm{e}^{i\xi x /d} \mathrm{e}^{-\frac{i\hbar\xi^2 (t-t')}{2md^2}},
\label{eq:Psi_1_Xi}
\end{align}
where $Q(t,z)$ is defined in Eq.~\eqref{eq:Q} and
\begin{align}
\Xi_l (t, t', x, y, \xi) &= \sigma^l \mathrm{e}^{-\beta^2/(4\alpha)} \sum_{n=0}^{[l/2]} (-1)^n C_l^{2n} \frac{(2n-1)!!}{2^n} \sum_{k=0}^{[l/2]-n} C_{l-2n}^{2k} \frac{(2k-1)!!}{2^k} (4\hbar^2 \alpha)^{n+k} (iy + \beta \hbar)^{l-2n-2k} \nonumber \\
{}&=\sigma^l (iy + \beta \hbar)^l \mathrm{e}^{-\beta^2/(4\alpha)}.
\label{eq:Xi}
\end{align}
Here $\alpha$ and $\beta$ involve $t$, $t'$, $x$ and $\xi$ and are given by Eqs.~\eqref{eq:alpha_beta}. It turns out that only the term with $n=k=0$ contributes, which can be easily shown by fixing $j = n+k$ ($0\leqslant j \leqslant [l/2]$) and summing over $n = 0$, $1$, $\ldots j$. The latter sum vanishes unless $j=0$. Let us now represent Eq.~\eqref{eq:Psi_1_Xi} in the following final form:
\begin{equation}
\Psi^{(1)}(t,\mathbf{r}) = -\frac{i}{2\hbar d} \frac{1}{\pi^{7/4} \sqrt{l!}} \frac{\sigma^{3/2}}{[\sigma \sigma_\perp(t)]^{2l+2}} \, \bigg ( 1 - \frac{it}{t_\text{d}} \bigg )^{l+1} Q(t,z) \, \mathrm{exp} \bigg [ \! - \frac{\rho^2 ( 1 - it/t_\text{d} )}{2[\sigma_\perp(t)]^2} \bigg ] \int\limits_{-\infty}^{\infty} \! d\xi \, \mathcal{I}_l (t, x, y, \xi) e\hat{\phi}(\hbar \xi /d) \mathrm{e}^{i\xi x /d},
\label{eq:Psi_1_final}
\end{equation}
where
\begin{equation}
\mathcal{I}_l (t, x, y, \xi) = \int \limits_0^t \! d\tau \, \mathrm{exp} \bigg [ -\frac{\xi^2}{4\alpha m^2 d^2} \, \tau^2 + \frac{\xi}{2md} \bigg ( \frac{x}{\alpha \hbar} - \frac{i\hbar \xi}{d} \bigg ) \tau \bigg ] \bigg ( x + iy - \frac{\hbar \xi}{md} \, \tau \bigg )^l.
\label{eq:Il}
\end{equation}
The integrations over $\xi$ and $\tau$ in Eqs.~\eqref{eq:Psi_1_final} and \eqref{eq:Il} will be performed numerically. Although the $z$ dependence of $\Psi^{(1)}(t,\mathbf{r})$ is given by the same profile $Q(t,z)$ as that revealed in the zeroth-order wave function~\eqref{eq:Psi_0}, the first-order correction~\eqref{eq:Psi_1_final} is highly nontrivial as a function of the transverse coordinates $x$ and $y$. \AK{Namely, it is already clear that the transverse part in Eq.~\eqref{eq:psi0}, which is given by $(x+iy)^l \mathrm{exp} (-\sigma^2 \rho^2/2)$, is nontrivially affected by the external field for $t>0$. For instance, the resulting probability density is no longer $\varphi$ independent as will be evidently demonstrated in what follows.}
\end{widetext}

Let us now discuss how one can perform the principal-value integration over $\xi$ given the explicit form of $\hat{\phi}$ in Eq.~\eqref{eq:phi_P}. We will choose the potential profile $f(\xi)$ in the following form
\begin{equation}
f(\xi) = \begin{dcases}
1, & \xi \geqslant 1, \\
\sin\, (\pi\xi/2), & -1 < \xi <1, \\
-1, & \xi \leqslant -1.
\end{dcases}
\end{equation}
In this case,
\begin{equation}
\hat{g}(\eta)=\frac{2i}{\eta}\frac{4\eta^2-\pi^2+\pi^2\cos{\eta}}{4\eta^2-\pi^2}.
\end{equation}
This is a smooth odd function which exhibits a $1/\eta$ behavior for large $|\eta|$. It vanishes rather slowly since the function $g(\xi)$ has discontinuities at $\xi = \pm 1$ no matter how smooth the profile $f(\xi)$ is. We now obtain
\begin{equation}
e\hat{\phi}(\hbar \xi /d) = -eE_0 d^2 \mathrm{e}^{-ia\xi/d} \bigg [ \hat{g}(\xi) - 2 i \mathcal{P} \, \frac{1}{\xi} \bigg ],
\end{equation}
which is to be plugged into Eq.~\eqref{eq:Psi_1_final}. To evaluate an integral of the form
\begin{equation}
I = \int \limits_{-\infty}^{\infty} \! d\xi \, h(\xi) \bigg [ \hat{g}(\xi) - 2 i \mathcal{P} \, \frac{1}{\xi} \bigg ],
\end{equation}
one can use the following simple prescription:
\begin{equation}
I = 2i\pi^2 \int \limits_{-\infty}^{\infty} \! d\xi \, [h(\xi) - h(0)] \, \frac{\cos \xi}{\xi (4\xi^2 - \pi^2)}.
\end{equation}
This is an ordinary Riemann integral which does not require any regularizations. Moreover, in our setup, $h(\xi)$ rapidly vanishes for $|\xi| \to \infty$ due to the exponential factor in Eq.~\eqref{eq:Il}, so the numerical convergence for large $|\xi|$ is quite fast.


\section{Numerical results and discussion} \label{sec:results}

To provide a clear roadmap for the reader, this section is arranged in a strictly parallel way for the two setups~\eqref{eq:app_pot} and \eqref{eqn:profile}. We begin with the analysis of the fully-localized (delta-like) potential and then examine the more complex field configuration involving a transverse inhomogeneous electric field. Inside each subsection, we numerically investigate how the initial twisted electron wave packet~\eqref{eq:psi0} with a given OAM $l$ evolves under the influence of the different perturbations. We first study the general properties of the probability density $\lvert\Psi(\mathbf r,t)\rvert^{2}$ and then analyze in detail the effects of vortex splitting.

\begin{figure}[t]
    \centering
    \includegraphics[width=\linewidth]{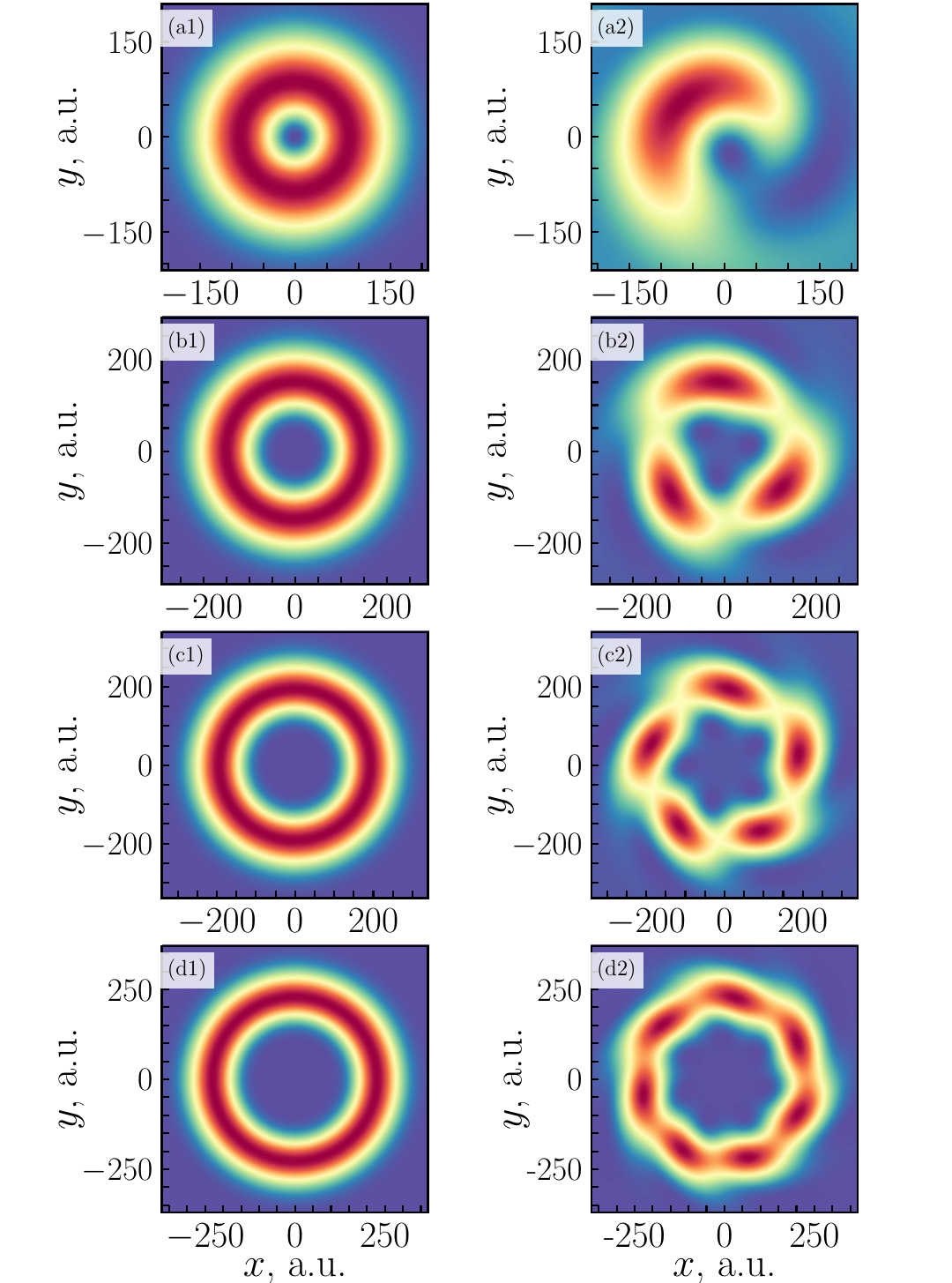}
\caption{Electron probability density for a twisted wave packet interacting with a $\delta$-like perturbation~\eqref{eq:app_pot}. Rows (a)-(d) correspond to odd OAM projections $l=1$, $3$, $5$, and $7$, respectively. Left column displays the unperturbed density $|\Psi^{(0)}|^{2}$; right column contains the plots for $|\Psi^{(0)}+\Psi^{(1)}|^{2}$. When the off-axis $\delta$-like potential at $\rho_{0}=10$~a.u. is included [panels (a2)-(d2)], the ring structure acquires an $l$-fold azimuthal modulation consisting of $l$ maxima (red). The coupling strengths are $\lambda=30$~a.u., $0.2$~a.u., $0.007$~a.u., and $0.0002$~a.u. for the four $l$ values, respectively. The other parameters are the following: the electron energy is $\bar{p}^2/(2m) = 2$~keV, $\sigma = 0.02$~a.u., and $t = 3500$~a.u. \AK{The average speed of the electron in units of the spped of light amounts to $\bar{v}/c \approx 0.09$, which justifies our nonrelativistic treatment of the problem.}}
    \label{fig:app_1}
\end{figure}
\begin{figure}[t]
    \centering
    \includegraphics[width=\linewidth]{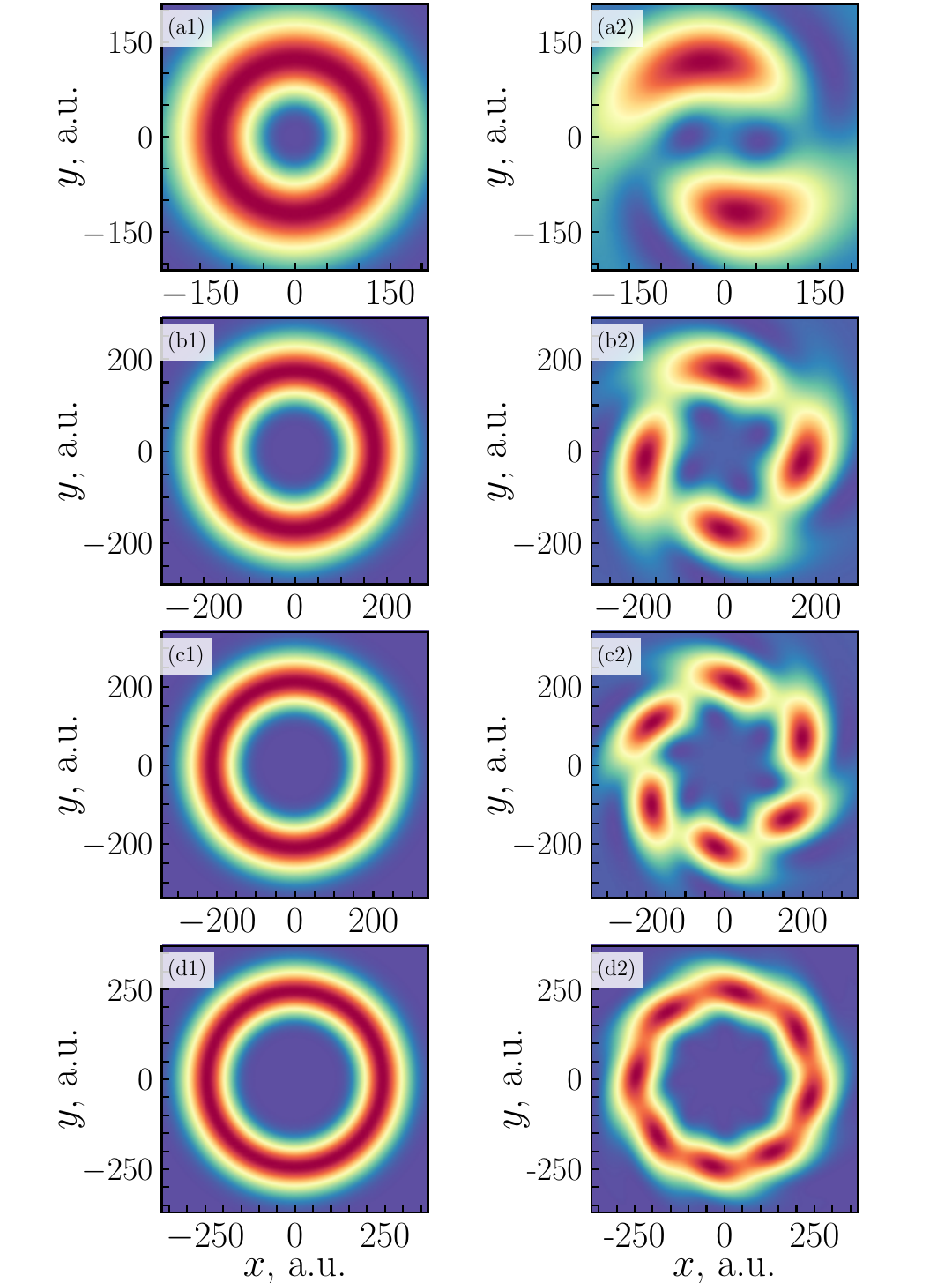}
\caption{The same as in Fig.~\ref{fig:app_1} for even values of the OAM projection: $l=2$, $4$, $6$, and $8$ [rows (a)--(d)]. The coupling strengths are $\lambda=3$~a.u., $0.045$~a.u., $0.002$~a.u., and $4\times10^{-5}$~a.u., respectively. The rest parameters are the same as in Fig.~\ref{fig:app_1}.}
    \label{fig:app_2}
\end{figure}
    
\subsection{Fully-localized perturbation} 
 \label{sec:subsection_a}
 
\subsubsection{Evolution of the density profile}

We first examine how a fully-localized external potential~\eqref{eq:app_pot} affects the transverse structure of the initial LG wave packet. To this end, we numerically evaluate Eq.~\eqref{eq:app_Psi_1}. In Fig.~\ref{fig:app_1} we summarize our data for the unperturbed electron probability density $|\Psi^{(0)}|^{2}$ and the total density $|\Psi^{(0)}+\Psi^{(1)}|^{2}$ for odd values of the OAM projection: $l=1$, $3$, $5$, and $7$. In the absence of the perturbation (left column), each state forms a ring whose radius depends on $l$, exactly as predicted by the free particle evolution. Introducing a $\delta$-like potential at $\rho_{0}=10$~a.u. on the $x$ axis ($z_{0}=0$) violates the perfect cylindrical symmetry and substantially changes the ring structure: the localized perturbation imprints a pronounced azimuthal modulation consisting of $l$ maxima along the original ring that form a regular polygon. To keep the ring‐segmentation effect equally discernible for every $l$, we deliberately scale the coupling constant $\lambda$ with $l$. In Fig.~\ref{fig:app_2} we present the results of analogous calculations in the case of even OAM projections: $l=2$, $4$, $6$, and $8$. We reveal a similar behavior, now with $2$, $4$, $6$, and $8$ density peaks. The density patterns are symmetric with respect to the reflection $x \to -x$, $y \to -y$.

\begin{figure}[t]
    \centering
    \includegraphics[width=\linewidth]{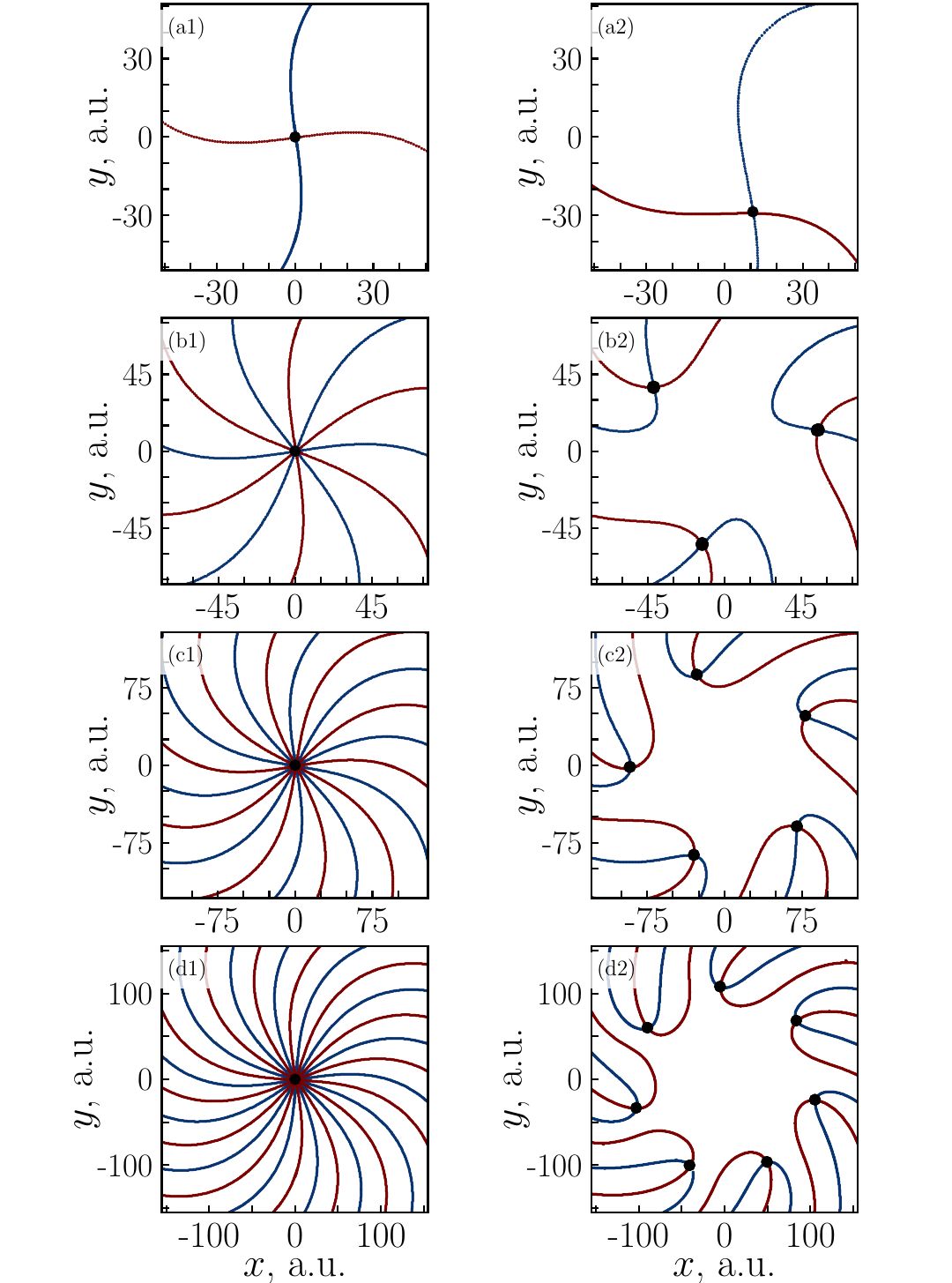}
    \caption{Nodal structure of the twisted electron packet perturbed by the localized potential~\eqref{eq:app_pot}. Blue (red) curves trace the zero level of $\mathrm{Re}\,\Psi$ ($\mathrm{Im}\,\Psi$); their intersections (black dots) mark density zeros. Rows (a)--(d) correspond to odd orbital charges $l=1$, $3$, $5$, and $7$, respectively.  Left column shows the unperturbed state $\Psi^{(0)}$, where a single $l$-fold node resides at the origin. Right column depicts the full wave function $\Psi^{(0)}+\Psi^{(1)}$. The $\delta$-like potential located at $\rho_{0}=10$~a.u. splits the central node into $l$ first-order vortices whose positions form a regular polygon: one off-axis zero for $l=1$, three for $l=3$ and so on. The coupling strength $\lambda$ and other parameters are the same as in Fig.~\ref{fig:app_1}.}
    \label{fig:app_3}
\end{figure}
\begin{figure}[t]
    \centering
    \includegraphics[width=\linewidth]{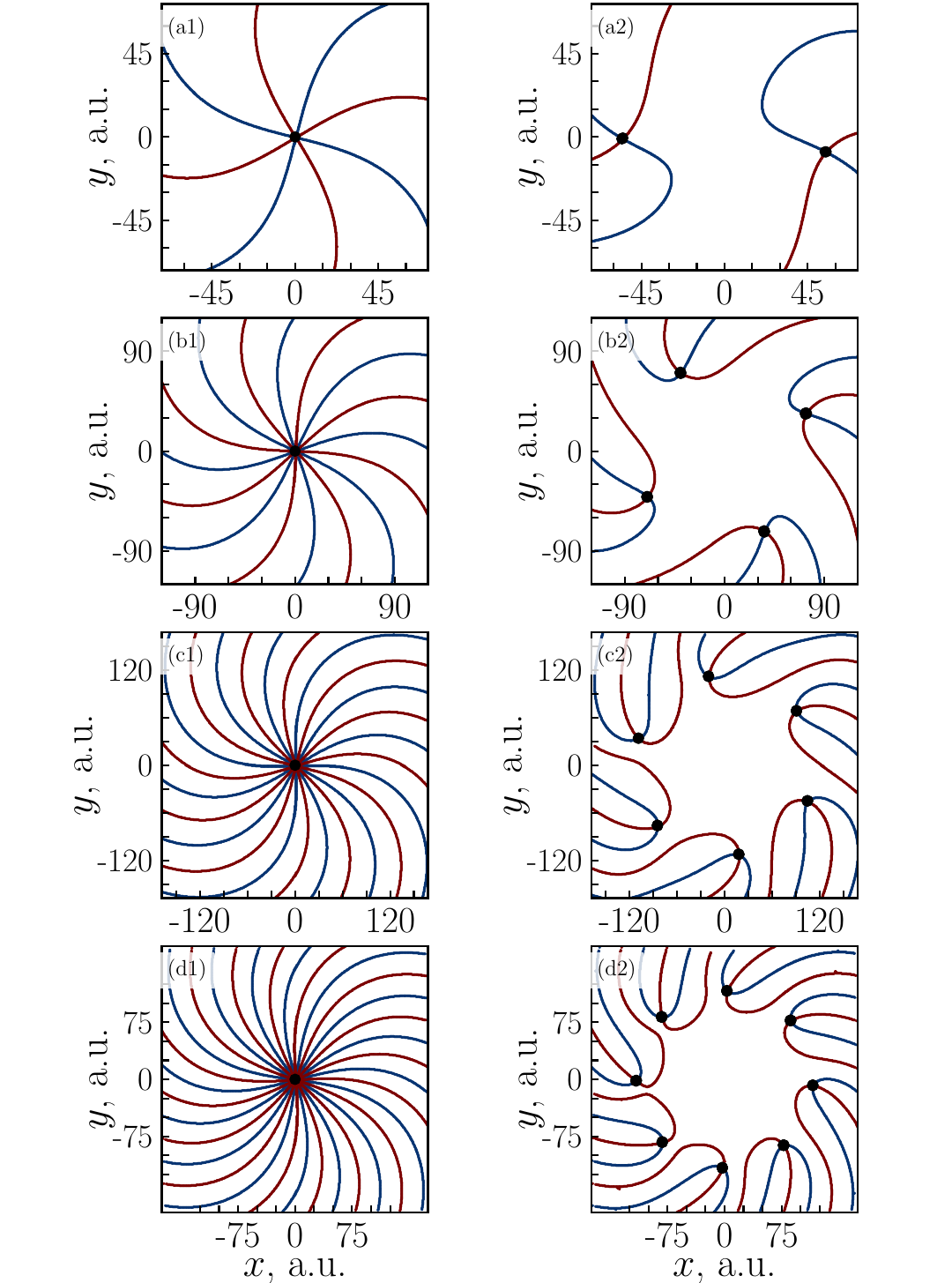}
    \caption{The same as in Fig.~\ref{fig:app_3} for even values of the OAM projection: $l=2$, $4$, $6$, and $8$. The coupling strength $\lambda$ and other parameters are the same as in Fig.~\ref{fig:app_2}.}
    \label{fig:app_4}
\end{figure}

\subsubsection{Splitting of the density zeros}

It is a well-known effect in the context of optical beams with high topological numbers $l$ that the initial zero of the local intensity at $x=y=0$ with multiplicity $l$ can split into several nondegenerate zeros if the beam gains additional corrections with lower topological charges (see, e.g., Refs.~\cite{KruglovLogvinVolkov1992,BaranovaZeldovich1981,Soskin2001,Rosanov2023}). In a purely linear and homogeneous medium, the initial singularity splits into multiple single‐charge vortices, typically arranged on a circle whose radius grows along the propagation direction. This phenomenon can be viewed as an \textit{instability} of higher‐order vortices against symmetry‐breaking perturbations: because topological charges greater than unity are effectively superpositions of single‐charge vortices, any perturbation that breaks perfect azimuthal symmetry can induce their spatial separation.

In the case of twisted electrons, the initial state~\eqref{eq:psi0} contains $\rho^l$ which indicates the presence of a degenerate zero at the origin of the transverse $xy$ plane. In this section, we will demonstrate that, similarly to the optical effect, this zero, in fact, fully splits into separate first‐order zeros due to the interaction with the external symmetry-breaking electric background. Although the physical system here is governed by the nonrelativistic Schr\"odinger equation rather than the paraxial wave equation of classical optics, the mathematical analogies between the two formalisms lead to closely related outcomes.

In Fig.~\ref{fig:app_3} we present the zero contours of $\mathrm{Re}\,\Psi$ and $\mathrm{Im}\,\Psi$ for $l = 1$, $3$, $5$, and $7$. In the left column we, the unperturbed wave function contains a degenerate singularity at $x=y=0$ representing a common intersection point of $2l$ contour lines. The localized external perturbation breaks the azimuthal symmetry, so that each high-order vortex splits into $l$ first-order vortices that migrate off axis and settle at equal azimuthal angles (right column in Fig.~\ref{fig:app_3}). in Fig.~\ref{fig:app_4} the same analysis is performed for even values of $l$. We observe that for any given $l$, the new (nondegenerate) zeros lie on a circle centered at the origin and form a regular polygon with $l$ vertices. Note that no residual central zero survives, indicating a full redistribution of the singularities.

Our results demonstrate that the well-known instability of higher-order optical vortices has an evident analog in the context of electron twisted states. Next, we will turn to the analysis of the transverse external field~\eqref{eqn:profile}, which represents a more involved scenario. We will show that the main qualitative patterns remain in this case, whereas the structure of the resulting wave packet turns out to be not that regular.

\subsection{Transverse inhomogeneous field}\label{sec:subsection_b}

\subsubsection{Evolution of the density profiles}

In this section, we study the influence of the inhomogeneous external electric field along the $x$ axis, Eq.~\eqref{eqn:profile}. To obtain the first-order correction $\Psi^{(1)}$ to the electron wave function, we numerically compute Eq.~\eqref{eq:Psi_1_final}.

\begin{figure}[t]
    \centering
    \includegraphics[width=1\linewidth]{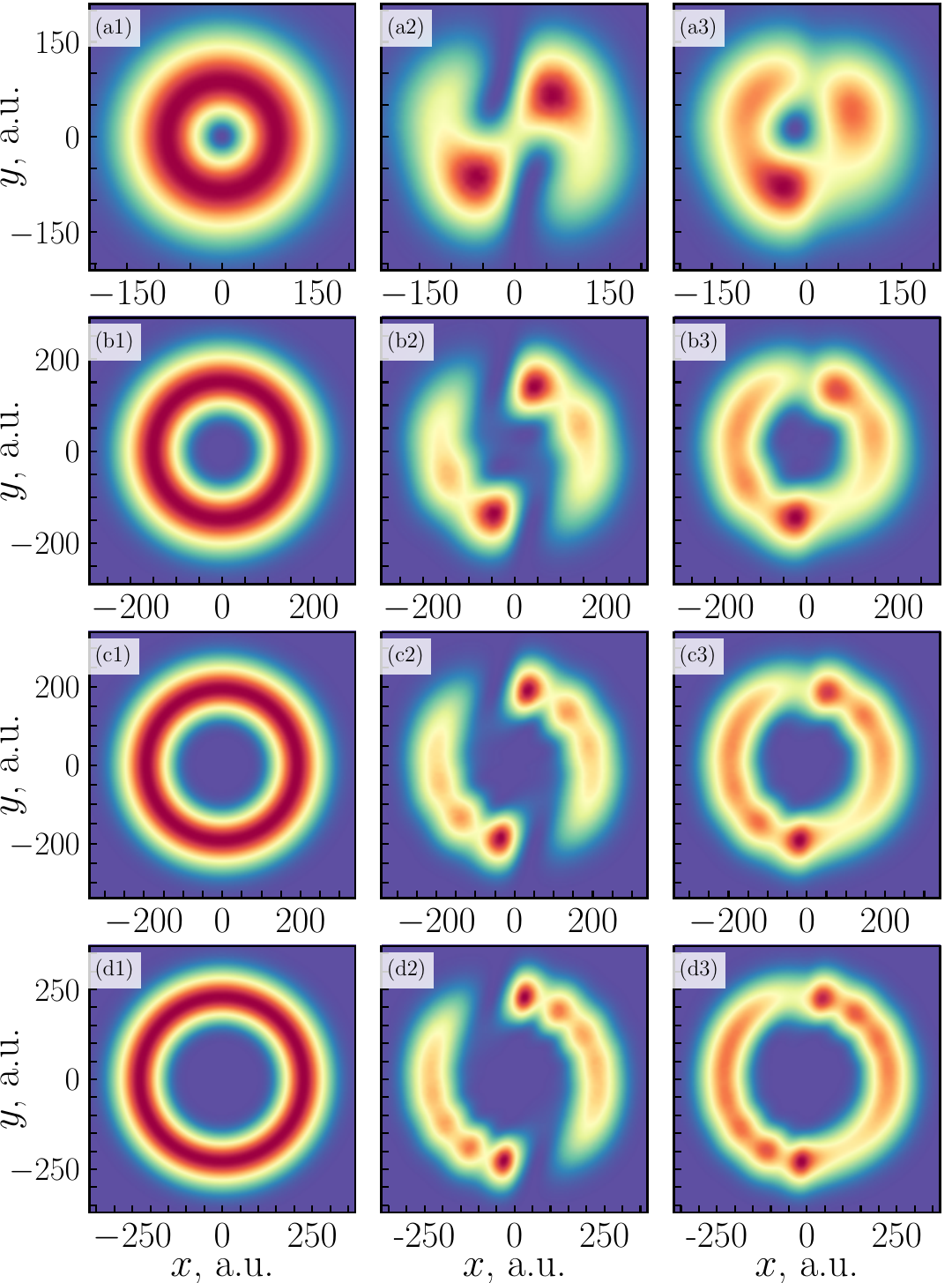}
    \caption{Electron probability density for a twisted wave packet under the influence of an inhomogeneous external electric field~\eqref{eqn:profile}. The rows correspond to different OAM values $l = 1$, $3$, $5$, and $7$ (from top to bottom). The first column shows the unperturbed density $|\Psi^{(0)}|^2$, the second column contains the first-order correction $|\Psi^{(1)}|^2$, and the third column displays the total density $|\Psi^{(0)}+\Psi^{(1)}|^2$. The relatively large values of the probability density are depicted in red. The graphs illustrate that the external perturbation significantly modifies the vortex structure, particularly affecting the maxima along the initial ring. The parameters are the following: the energy is $\bar{p}^2/(2m) = 2$~keV, $\sigma = 0.02$~a.u., $t = 3500$~a.u., $d=10$~a.u., $a=0$, and $E_0=1.0 \times 10^7$~V/m.}
    \label{fig:enter-label}
\end{figure}

Figure~\ref{fig:enter-label} presents the electron probability density profiles for four different values of $l$, from top to bottom: $l=1$, $3$, $5$, and $7$. In each row, the left panel shows $|\Psi^{(0)}|^2$, i.e. the unperturbed density; middle panel corresponds to $|\Psi^{(1)}|^2$, illustrating the first-order correction alone; and the right panel demonstrates the total density $|\Psi^{(0)}+\Psi^{(1)}|^2$. As was already seen, the unperturbed wave packet exhibits a simple ring‐like structure with a pronounced maximum along the radial direction independent of the polar angle $\varphi$. The full wave function $\Psi^{(0)}+\Psi^{(1)}$ possesses a nontrivial azimuthal dependence. In contrast to the symmetric polygonal splitting revealed in the case of the $\delta$-like perturbation in Sec.~\ref{sec:subsection_a}, the resulting density profile is not that regular. We focus on the case $a=0$, where the first-order correction has the following {\it exact} symmetry: $\Psi^{(1)}(t,-x,-y,z) = (-1)^{l+1} \Psi^{(1)}(t,\mathbf{r})$, which can be directly seen from Eq.~\eqref{eq:Psi_1_final} (one has to perform the variable change $\xi \to -\xi$ in the integral). The probability density is therefore also symmetric with respect to the above reflection. 

In Fig.~\ref{fig:enter-label} we observe that once the inhomogeneous electric field is included, the radial rings develop maxima and minima along the azimuthal direction. The first‐order correction term $\Psi^{(1)}$ distorts the original cylindrical symmetry of the unperturbed LG mode, effectively creating localized ``hot spots'' on the ring. Furthermore, for larger $l$, more maxima and minima appear around the circumference. This arises because higher-$l$ modes have a stronger inherent azimuthal phase dependence, so even a relatively weak field inhomogeneity can mix the neighboring OAM modes, leading to additional angular modulation in the total density. Let us note, while the number of azimuthal modulations still increases with $l$, the potential~\eqref{eqn:profile} blurs the peaks more than the point-source perturbation, so the correspondence between $l$ and the count of density hotspots along the ring becomes noticeably less distinct. We underline here that although our calculations are limited to the first order of perturbation theory, the qualitative patterns revealed are quite universal. The external-field amplitude represents a trivial prefactor in the first-order correction and chosen to be $E_0 = 1.0 \times 10^7$~V/m only for the presentation reasons. For lower amplitudes, the structure of the wave packet will be modified more weakly from the quantitative perspective, while for larger $E_0$ one will need to take into account the higher-order contributions. Nevertheless, the qualitative effect of symmetry breaking will remain.

\begin{figure}[t]
    \centering
\includegraphics[width=1\linewidth]{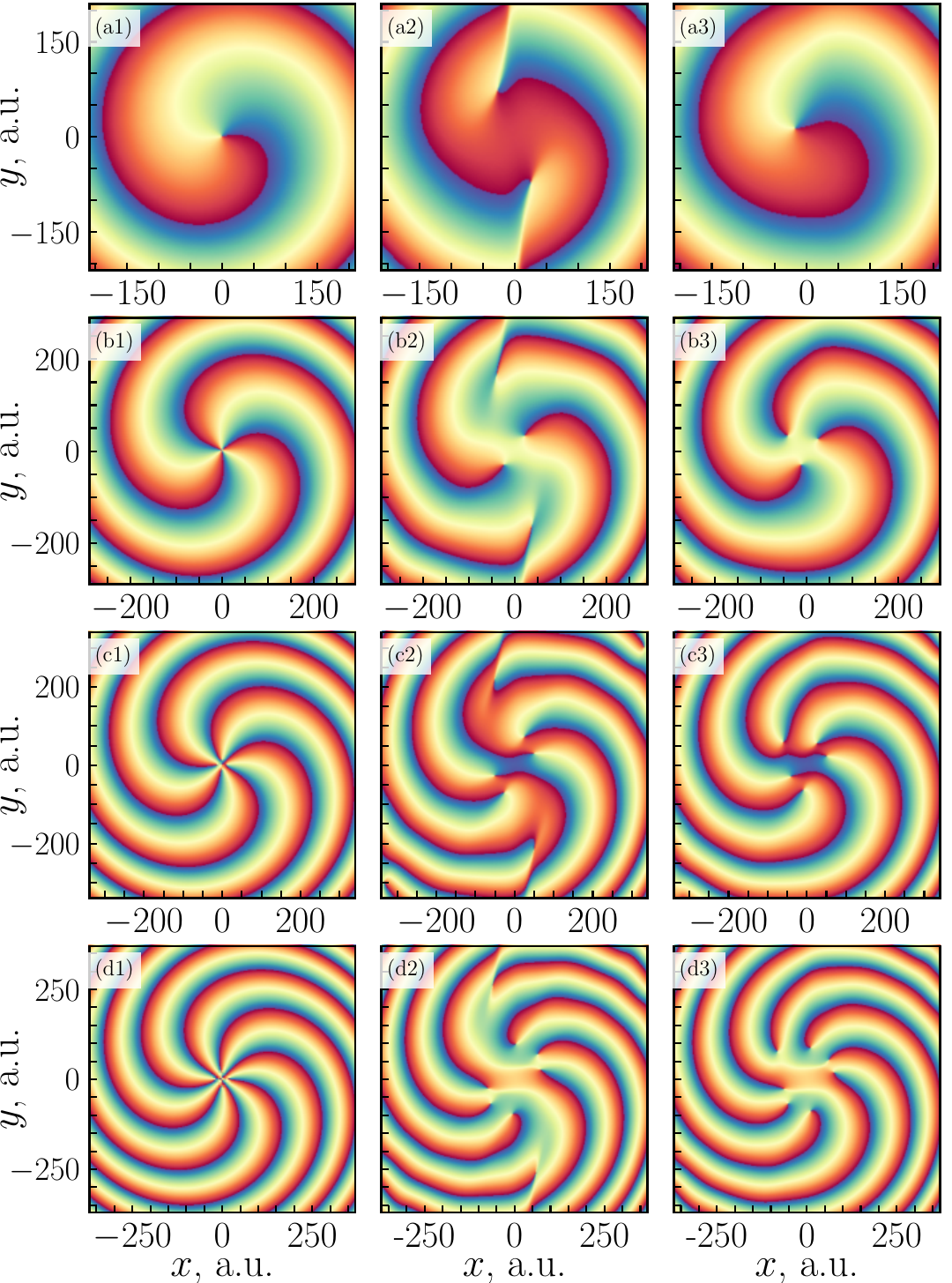}
    \caption{\AK{Phase} of the twisted electron's wave function under the influence of an inhomogeneous external electric field~\eqref{eqn:profile}. The rows correspond to different OAM values: $l = 1$, $3$, $5$, and $7$ (from top to bottom). The first column shows the unperturbed phase $\mathrm{Arg}\,\Psi^{(0)}$, the second column displays the first-order correction $\mathrm{Arg}\,\Psi^{(1)}$ \AK{(illustrative; vortices here need not coincide with those of the total wave function)}, and the third column corresponds to the \AK{phase} of the total wave function  $\mathrm{Arg} \, [\Psi^{(0)}+\Psi^{(1)} ]$. Positive (negative) values are displayed in red (blue). The plots demonstrate that the external perturbation significantly modifies the phase structure and wavefront shape. The parameters of the system are the same as in Fig.~\ref{fig:enter-label}.
    }
    \label{fig:enter-label2}
\end{figure}

In Fig.~\ref{fig:enter-label2} we show the \AK{phase} of the wave function for the same four values of $l$, \AK{i.e., we display the argument of the corresponding complex numbers, $\mathrm{Arg} \, \Psi = \mathrm{Arg} \, (|\Psi|\mathrm{e}^{i\chi}) = \chi$, which takes the values from $-\pi$ to $\pi$. Phase singularities are points where the wave function vanishes, so that the phase is undefined.} Analogous to Fig.~\ref{fig:enter-label}, the first column of each row displays $\mathrm{Arg}\,\Psi^{(0)}$, the second column corresponds to $\mathrm{Arg}\,\Psi^{(1)}$, and the third column shows $\mathrm{Arg}\,[\Psi^{(0)}+\Psi^{(1)}]$. Whereas the density plots in Fig.~\ref{fig:enter-label} emphasize how the probability distribution is reshaped, 
the \AK{phase} plots more directly reveal the wavefront distortions. The unperturbed wave function has a spiral-type profile of a helically phased LG beam. However, the presence of the external field makes the spiral arms uneven, creating additional azimuthal modulations. Again, with increasing $l$, the spiral pattern becomes more complex and the external-field effects lead to a more involved structure of the resulting state. \AK{We should also note, that the middle column displays the phase of $\Psi^{(1)}$; vortices seen there are not, in general, the zeros of the physical state. Our vortex analysis is based on the rightmost column showing the full field.}

Together, Figs.~\ref{fig:enter-label} and \ref{fig:enter-label2} highlight the important conclusion of our study: twisted electron beams carrying higher orbital angular momentum are particularly sensitive to the influence of inhomogeneous fields, which can induce pronounced deformations in both the probability density and the phase structure. These results underscore the importance of carefully controlling ambient fields in experiments.

\subsubsection{Splitting of the density zeros}
\label{subsec:splitting}

In this subsection, we perform the analysis of the splitting effect in the case of the $x$-dependent inhomogeneous electric field. However, in contrast to both the ideal optical scenario and the highly symmetric splitting produced by the $\delta$-like potential, the present field introduces additional kinetic- and potential-dependent contributions that disturb the perfect geometry of the emerging single-charge vortices. As a result, the polygonal arrangement of nodes becomes noticeably less regular.

\begin{figure}[t]
    \centering
    \includegraphics[width=\linewidth]{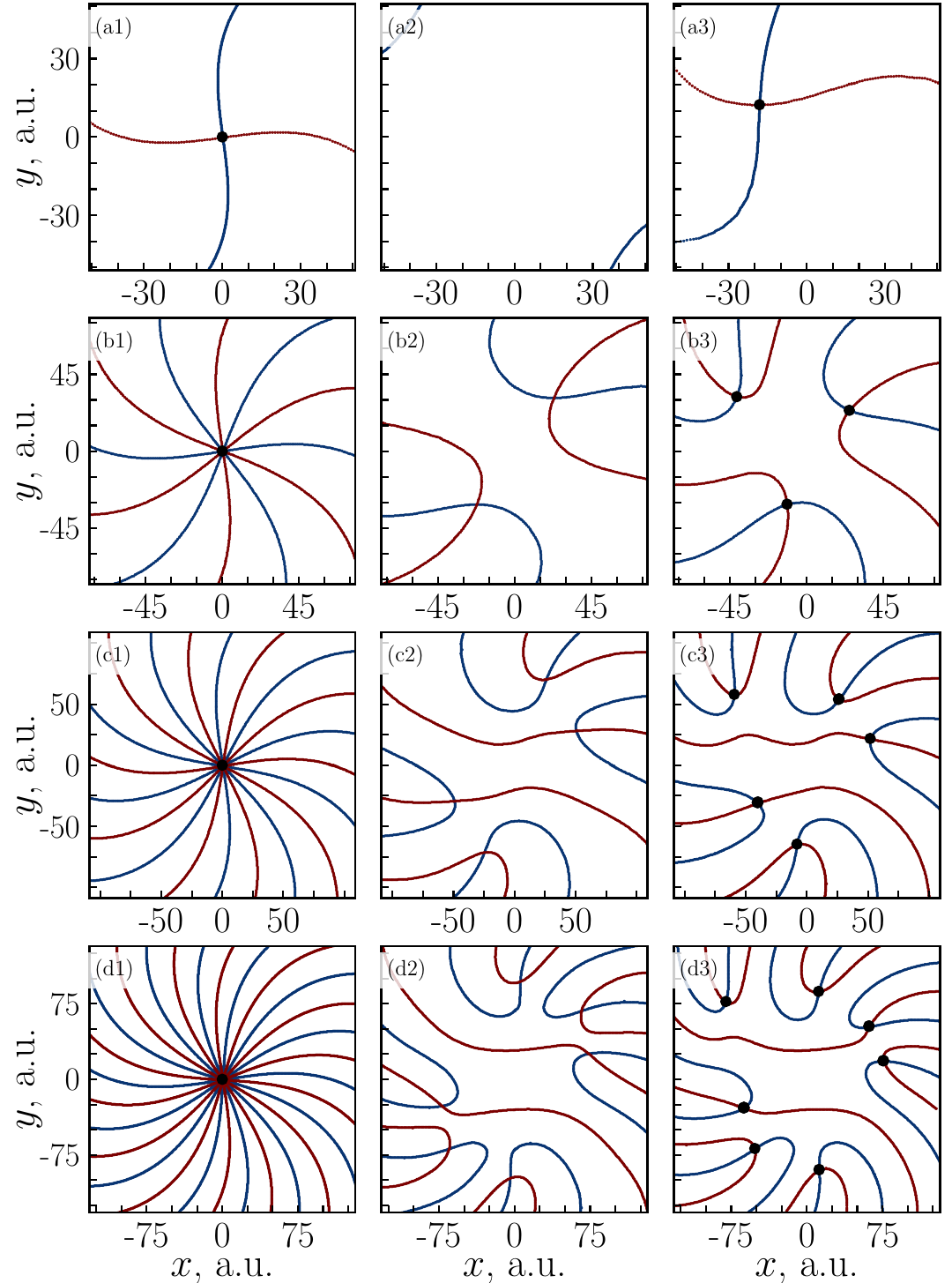}
    \caption{Zero‐level contour lines of the real (blue) and imaginary (red) parts of the wave function for a twisted electron wave packet with OAM $l = 1$, $3$, $5$, and $7$ [rows (a)--(d)]. The three columns show, from left to right, the unperturbed component $\Psi^{(0)}$, the first‐order correction $\Psi^{(1)}$, and the full wave function $\Psi^{(0)}+\Psi^{(1)}$. \AK{Black dots are shown only in the rightmost column and mark intersections of the real and imaginary nodal lines, i.e., the density zeros (phase singularities) of the full state. We deliberately do not mark intersections in the middle column because zeros of $\Psi^{(1)}$ alone are not, in general, the physically relevant vortices of the total wave function.} While $\Psi^{(0)}$ features a single $l$-fold zero at the origin, the perturbed and total wave functions exhibit splitting into multiple singly‐charged vortices whose spatial arrangement becomes progressively less regular with increasing $l$. The system parameters are the same as in Figs.~\ref{fig:enter-label} and \ref{fig:enter-label2}.
    }
    \label{fig:zeros}
\end{figure}

\begin{figure}[t]
    \centering
    \includegraphics[width=\linewidth]{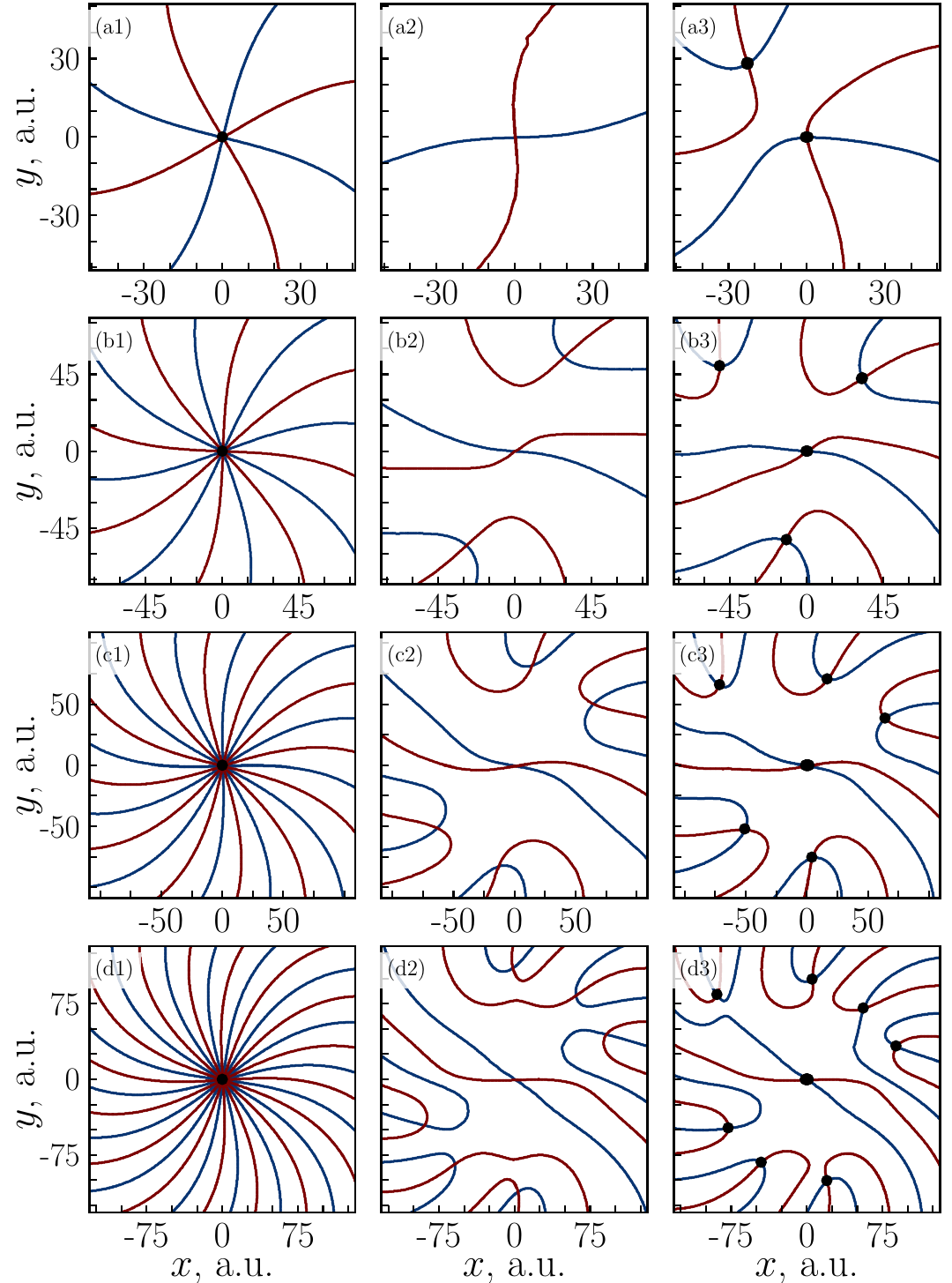 }
    \caption{The same as in Fig.~\ref{fig:zeros} for even values of the OAM projection: $l=2$, $4$, $6$, and $8$ for rows (a)--(d), respectively.}
    \label{fig:zeros2468}
\end{figure}

In Fig.~\ref{fig:zeros} we show the zero‐level contours of the real (blue) and imaginary (red) parts of the wave function for odd orbital angular momentum $l=1$, $3$, $5$, and $7$ [rows (a)--(d)]. In each row, the left panel displays the nodal lines of the unperturbed mode $\Psi^{(0)}$, the middle panel corresponds to those of the first‐order correction $\Psi^{(1)}$, and the right panel shows those of the full wave function $\Psi^{(0)}+\Psi^{(1)}$. The black dots mark the intersections of the real and imaginary nodal lines, i.e., the locations of singly‐charged vortices or true zeros of the electron wave packet. One clearly sees that the single $l$‐fold zero at the center of $\Psi^{(0)}$ is replaced, upon OAM admixture, by $l$ off‐axis vortices whose azimuthal positions are not perfectly regular but reflect the combined effects of wave‐packet kinetics and the inhomogeneous field.

Figure~\ref{fig:zeros2468} presents the analogous nodal‐line maps for even charges $l=2$, $4$, $6$, and $8$. Again, the three columns correspond to $\Psi^{(0)}$, $\Psi^{(1)}$, and $\Psi^{(0)}+\Psi^{(1)}$, respectively, and black dots denote the individual singularities in the full wave function. As before, the central $l$‐fold zeros split into $l$ separate vortices, but now the zero at $x=y=0$ remains becoming a nondegenerate one. This is an exact consequence of the symmetry property $\Psi^{(1)}(t,-x,-y,z) = (-1)^{l+1} \Psi^{(1)}(t,\mathbf{r})$, leading to $\Psi^{(1)}(t, 0, 0, z) = 0$ for even $l$. This indicates that the splitting pattern strongly depends on the parity of the original topological charge as well as on the parity properties of the perturbation. Here we again underline the universal character of the splitting effect. For instance, since the first-order correction does not vanish at the origin $x=y=0$ for odd $l$, the zeros of the total wave function $\Psi^{(0)}+\Psi^{(1)}$ inevitably split no matter what the external-field amplitude is chosen. The actual value of $E_0$ governs only the quantitative characteristics of the splitting pattern.

If the interaction with the external electric field has a finite duration $0\leqslant t \leqslant \tau$, then once the external field has been switched off ($t>\tau$), each OAM component in the combination $\Psi^{(0)}+\Psi^{(1)}$ evolves freely and therefore retains its own time-dependent width $\sigma_\perp(t) = \sigma^{-1}(1+t^{2}/t_{\mathrm d}^{2})^{1/2}$ [see Eq.~(\ref{eq:Psi_0})]. Accordingly, a zero revealed at transverse radius $\rho_{k}$ at the end of the interaction ($t=\tau$) continues moving away from the origin, and for $t \gg t_\text{d}$ it drifts \emph{linearly} with $t$: $\rho_{k} (t) \sim \rho_{k} (t/\mathrm{max}\{t_\text{d}, \tau\})$. As a result, the pairwise distance between any two of the zeros also grows linearly with $t$, so the structures uncovered above expand self-similarly.

Overall, these results serve as a bridge between the extensive knowledge of optical vortex splitting in linear media and the emergent field of twisted electron beams. They also suggest that the stability of high‐order electron vortices may be an even subtler issue, as additional factors (e.g., potential gradients and relativistic corrections at higher energies) can further influence or catalyze the splitting process. 

\section{Conclusion} \label{sec:conclusion}

\AK{In this paper, we analyzed the evolution of a twisted electron Laguerre–Gaussian wave packet in a static, inhomogeneous electric field that breaks axial symmetry about the $z$ axis --- the axis of initial electron density symmetry. We obtained closed-form first-order expressions  for the field-induced correction to  the wave function and used them to track how symmetry breaking reshapes the transverse density and phase. In particular, we identified the mechanism by which the initial $l$-fold zero at the beam center splits into multiple singly charged vortices, and we showed that for a realizable one-dimensional profile $E_x(x)$ the survival or removal of the central zero depends on the parity of $l$. We also established simple scaling laws for the post-interaction evolution: once the field is switched off, the vortex expands self-similarly with a radius that grows linearly in time.

Beyond reproducing the broadly expected tendency of high-order vortices to split under symmetry breaking, these results provide a compact framework tailored to {\it twisted electrons} in electrostatic inhomogeneities. The formulas make direct contact with observables: ring segmentation and azimuthal modulation in real-space density, and maps of phase singularities in electron holography. Together with a practical field geometry [parallel plates producing $E_x(x)$], this positions twisted electrons as sensitive probes of weak transverse field gradients.

Looking ahead, several directions are natural. (i) Go beyond the weak-coupling regime to quantify nonlinear mixing of OAM channels and possible vortex–vortex interactions under stronger fields. (ii) Incorporate spin and relativistic corrections within the Dirac formalism in regimes where they cease to be negligible. (iii) Benchmark the predicted patterns against phase-resolved electron imaging and OAM spectroscopy, and explore the inverse use of our expressions to reconstruct transverse field profiles from measured vortex patterns. We expect this framework to serve as a baseline for quantitative beam diagnostics and as a bridge between singular optics and electron–matter-wave control in realistic electrostatic environments.}


\begin{acknowledgments}

The study was funded by the Russian Science Foundation, projects No.~23-12-00012 (the case of electron wave packets) and No.~21-72-30020-P (discussion of optical analogs). The numerical calculations in Sec.~\ref{sec:results} were supported by the Icelandic Research Fund (Ranns\'oknasj\'oður, Grant No.~2410550).

\end{acknowledgments}


\appendix*

\section{Wave packet dynamics in a homogeneous electric field}\label{sec:app_uniform_field}

Here we will demonstrate that the interaction with a spatially homogeneous electric field leads only to a trivial uniform shift of the probability density. We will not invoke here perturbation theory and explicitly show that this result is valid to all orders.

Let us consider a general case where the electric field has an arbitrary temporal dependence:
\begin{equation}
\phi(t, x) = - E(t) x.
\end{equation}
To get rid of the $x$ dependence of the Hamiltonian, we perform a gauge transformation
\begin{equation}
\Psi (t, \mathbf{r}) = \mathrm{e}^{-ieA(t)x/\hbar} \, \Psi_1 (t, \mathbf{r}),
\end{equation}
where the $x$ projection of the vector potential is given by
\begin{equation}
A(t) = - \int \limits_0^{t} \! E(t') dt'.
\end{equation}
The Schr\"odinger equation in terms of the yet-unknown function $\Psi_1 (t, \mathbf{r})$ reads
\begin{equation}
i \hbar \partial_t \Psi_1 (t,\mathbf{r}) = -\frac{\hbar^2}{2m} \bigg \{\Delta_{yz} + \Big[ \partial_x - \frac{ie}{\hbar} A(t) \Big ]^2 \bigg \} \Psi_1 (t,\mathbf{r}).
\end{equation}
Next, we absorb the $A^2$ term by means of the following phase factor:
\begin{equation}
\Psi_1 (t,\mathbf{r}) = \mathrm{exp} \Bigg [ - \frac{ie^2}{2m\hbar} \int \limits_0^t \! A^2 (t') dt' \Bigg ] \, \Psi_2 (t,\mathbf{r}).
\end{equation}
Then the function $\Psi_2 (t, \mathbf{r})$ obeys
\begin{equation}
i \hbar \partial_t \Psi_2 (t,\mathbf{r}) = \bigg [ -\frac{\hbar^2}{2m} \Delta + \frac{ie\hbar}{m} A(t) \partial_x \bigg ]  \Psi_2 (t,\mathbf{r}).
\end{equation}
If, at $t=0$, a classical electron at rest starts to interact with the external field $E(t)$, then by the time instant $t$ it travels a distance
\begin{equation}
s(t) = - \frac{e}{m} \int \limits_0^t \! A (t') dt'.
\end{equation}
This suggests the following substitution:
\begin{equation}
\Psi_2 (t,\mathbf{r}) = \Psi_3 (t, x - s(t), y, z).
\end{equation}
We find now that $\Psi_3 (t, \mathbf{r})$ satisfies a free-particle Schr\"odinger equation, i.e., it does not involve the external field at all. Furthermore, at the initial time instant $t=0$, the function $\Psi_3 (0, \mathbf{r})$ coincides with $\Psi (0, \mathbf{r}) = \Psi_0(\mathbf{r})$. This means that it is nothing but the unperturbed solution,
\begin{equation}
\Psi_3 (t, \mathbf{r}) = \Psi^{(0)} (t, \mathbf{r}).
\end{equation}
Finally, in terms of the probability density, we obtain
\begin{equation}
|\Psi (t, \mathbf{r})|^2 = |\Psi^{(0)} (t, x - s(t), y, z)|^2.
\end{equation}
Obviously, in the initial cylindrical coordinates $\rho$, $\varphi$, $z$, this shift violates the axial symmetry and gives rise to a combination of various projections $l$ of the angular momentum, but this does not affect the structure of the wave packet: for any given $t$, it is only necessary to properly shift the origin along the $x$ axis to recover the unperturbed probability density. Nevertheless, the effect of an {\it inhomogeneous} field cannot be described by such a shift as clearly seen in the graphs presented in Sec.~\ref{sec:results}.


\end{document}